# 0THE IMPACT AND INFLUENCE OF ACADEMIC GENEALOGIES

Bryan Briones[1], Ronald E. Mickens[2], and Charmayne Patterson[3]
[1] Atlanta University Center Robert W. Woodruff Library, Atlanta, GA 30314
[2] Department of Physics, Clark Atlanta University, Atlanta, GA 30314
[3] Department of History, Clark Atlanta University, Atlanta, GA 30314
[1] Corresponding author: bbriones@auctr.edu

We introduce the concept of an "academic genealogy" (AG) and illustrate how AG charts may be constructed by using Wikipedia as our reference. We then demonstrate how this methodology can be used by applying it to create the partial or full AG charts to two scientists: Paul A. Samuelson and Ronald E. Mickens. For Samuelson we find that not only did he have academic connections to some of the most creative economists of his time, but as his AG chart shows, Josiah W. Gibbs, his academic grandfather, was perhaps the most distinguished scientist to date, from North America. Similarly, the members of Mickens's AG can be traced back several centuries and include some of the most influential physicists and mathematicians of the 19th and 20th centuries. Thus, our major conclusion is that the AG of a scientist helps us to understand the path of their academic career, why they selected particular research topics, the impact on their discipline, and the success of their academic children.

## INTRODUCTION

An academic genealogy (AG) is a family tree of scholars based on mentoring relationships. In the sciences, these relationships are generally based on dissertation supervision. However, the AG may also be determined by starting with a subject's undergraduate mentor. A good introduction to this topic can be obtained from Wikipedia (Academic Genealogy, 2025).

In more detail, an AG is determined by finding out who was the mentor/advisor of the subject and continuing this process as far backward in time as available information will allow. If we plot these items in two dimensions, an AG chart is produced and may be used for a variety of purposes:

- to indicate the historical connections between the subject and earlier scientists and their research,
- provide insights into possible future paths for research on future topics,
- understand why a subject selected a particular topic to investigate for their research, and
- determine the social, academic, and research networks that shaped the career of a specific subject.

In essence, careful examination of the AGs of scientists who are connected by the same research agendas allows us to learn about the history of past achievement, and to determine who pioneered these achievements. For example, the roles played by knowledge of the history of science, particularly physics, is discussed in Reddington (2017) and Stanley (2016).

AGs have also been used to investigate and give insights into a number of related social science issues and behaviors. For instance:

- investigating the role of faculty advisors in the career pathways of their doctoral students (German, 2018),
- measuring the impact of medical AG on publication patterns and medical practice (Hirshman, 2016),
- determining the impact of an individual researcher on their discipline (Ridder, 2025),
- applying bibliometric methodologies to study scientific AGs (Ruihua, 2021),
- using AG to understand and predict academic success (Wuestman, 2020), and
- using AG to investigate the influence of coauthorship on researcher affiliation, research topic, productivity, and impact (Xie, 2022).

The construction of an AG can proceed in several ways. In the next section, we mainly focus on how this may be accomplished by using Wikipedia as a reference. However, one can also use information on the networked digital library of theses and dissertations (Dores, 2016).

Viewing AG within a wider context allows the introduction of the concept of an "invisible college" (Price, 1971; Crane, 1972; Paisley, 1972). Invisible colleges are groups of researchers who work in the same area, forming a network of informal contacts with one another. These groups generally have memberships of the order of 20-50 individuals. The following is a listing of the major characteristics of invisible colleges:

- Membership is not formalized but is strongly dependent on the acceptance of one's research by other members of the invisible college.
- The membership is generally not in the same or nearby geographical locations.
- The members' informal contacts and exchange of information, ideas, data, etc., can take place by using a wide variety of mechanisms. For example, in today's research environment, modes of information exchange generally take place via email, phone correspondence, social media platforms, or video teleconferencing.
- Members generally assume that researchers who are not in their circle are not doing research of value, as if to establish that their invisible college contains just the important players, making it an exclusive club of sorts.
- Members assume that other members can appreciate and understand each other's work and provide, if required, the appropriate interpretation.

- Members generally agree on what are the important issues, what methodologies should be used to resolve them, and can agree on when a suitable resolution has been obtained.

It should be noted that invisible colleges play very important roles in maintaining scientific progress and its administration. For example, members of the invisible college will generally:

- provide peer reviews of other members' manuscripts that are submitted to journals,
- sit on research panels to evaluate research funding proposals,
- serve on committees to make selections for high-value awards and honors,
- determine whether a given member is considered in elite professional organizations,
- influence the opportunities for other members to assume leadership of their discipline,
- provide career guidance, and
- shape particular scientific centered social interactions of members who are being mentioned by others in the invisible college.

Invisible colleges are dynamic, social, and scientific structures. Their membership changes with time and may dissolve when the initial critical issues or problems have either been "solved" or are no longer considered of any importance.

Finally, the invisible college can play a significant role in determining the value of any member's research and the resulting publications. This occurs because the member of the invisible college can encourage other members to cite other members' papers, and this can have both an immediate and significant impact on the value of these publications within the general scientific research community for that discipline.

With this background in hand, we can now state the main task of this paper, namely, to illustrate how the concepts of academic genealogy and the invisible college can be used to evaluate the careers of scientists. The next section provides a concise summary of our methodology. We then proceed to apply this methodology to first examine the career and influence of the economist, Paul A. Samuelson. This is followed by carrying out a similar analysis for physicist Ronald E. Mickens. Finally, we end the paper with a discussion of our results and a brief list of conclusions.

This paper is also an extension of a presentation made at the 2019 Annual Meeting of the Georgia Academy of Science, Section VI, History and Philosophy of Science (Mickens and Patterson, 2019).

## METHODOLOGY

The basic information on all of the persons discussed or presented in this work is obtained from the publicly available, online encyclopedia Wikipedia. For the natural sciences and related disciplines, their professional communities have worked very hard to ensure that data given by Wikipedia is accurate, up to date, and contains few errors or misinformation. These features generally hold for both technical and biographical entries given on their web pages.

Another important feature of Wikipedia's biographical entries is that they give for individuals holding a doctorate the name(s) of their advisor(s) and/or mentor(s), with a direct link to their Wikipedia entry. Consequently, it is rather easy to construct the academic genealogy of any given holder of a Ph.D. in the sciences.

In recent years, many science-based professional societies have become interested in providing data which allows their members and others the opportunity to generate or construct individual academic genealogies. Figure 1 gives a partial listing of several of these websites. Each year additional websites are being constructed for this purpose.

In this work, we do not provide the Wikipedia URLs for most of the persons who we place on the charts related to either the academic genealogies or the associated academic/research connections. This is largely done, as stated above, for the reasons that their information is easily available via their own Wikipedia entries.

The methodology used to construct the several academic genealogies is based on carrying out the following steps:

1) Go to Wikipedia (https://www.wikipedia.org) and search the subject by name.
2) Determine the subject's Ph.D. advisor.
3) Click the link to the subject's Ph.D. advisor and then determine that advisor's own Ph.D. advisor.
4) Continue this process for as far in the past as is needed or is possible.
5) From this information, construct the subject's academic genealogy chart.

Note that for the case of more than one Ph.D. advisor, the chart may become complicated as multi- (backward in time) trajectories exist.

Finally, Wikipedia provides a wealth of information for those who wish to explore the details of the life and career of the subject. Generally, on the right-hand side of the entry there is a panel of bulleted information as follows (not all of them may be available or known):

- Full name of the subject
- Date of birth and death
- Nationality
- Education and alma mater
- Concise summary of what they are known for
- Names of spouse(s) and child(ren)
- Research fields
- Institutional affiliations
- Title of thesis or dissertation
- Doctoral advisor

There might also be a link to an electronic copy of the subject's thesis or dissertation.

In this section, we apply the concept of academic genealogy to the case of Paul A. Samuelson, one of the most important and impactful economists of the twentieth century (Samuelson, 2007).

Figure 2 presents a truncated version of Samuelson's academic genealogy chart (AGC), covering his academic fathers and grandfathers and two of his academic children. However, using the procedure given in the section on Methodology, this chart may be expanded in either temporal direction.

Observe that Samuelson descended from a line of very distinguished late 18$^{th}$ and early 19$^{th}$ century economists. The AGC also indicates that two of his academic children, Lawrence Klein and Robert C. Merton, received the Nobel Prize in Economics, an honor bestowed to Samuelson in 1970. Samuelson used physical and mathematical concepts, reinterpreted, to raise the level of scientific analysis in economic theory (Samuelson, 2007). For example, he demonstrated mathematically that the existence of an equilibrium point in a supply-demand graph (explained in Pinkasovitch, 2024) did not necessarily imply that the equilibrium was stable.

Samuelson's work and writings influenced students worldwide and helped make MIT a premier center for the study and application of economics. Further extensions of this methodology produced what is now called "econophysics" (Farmer et al., 2025; Slanina, 2014; *Wikipedia*, 2025), the generalization of physical concepts from statistical and thermal physics for application to economical phenomena. One consequence of this approach was that economics could be made more logically rigorous. Thus, predictions could be made from the mathematical structures devised to model economic systems and, as a result, economic theories could be tested using actual data.

Figure 3 indicates that Samuelson's views and methodology regarding economics and how it should be thought about and approached were heavily influenced by Josiah Willard Gibbs (*Wikipedia*, 2025) via contact with Edwin Bidwell Wilson (*Wikipedia*, 2024). In a real sense, Gibbs is the most distinguished, impactful, and important American scientist to date; there are more than 23 concepts named after him (*Wikipedia*, 2025). Likewise, similar things can be said for Wilson (*Wikipedia*, 2024), who was a general polymath. Both individuals made major contributions to chemistry, mathematics, and physics. While Wison was a Ph.D. student of Gibbs at Yale University, Samuelson was a student, and later colleague, of Wilson at Harvard University and MIT. Both believed that complex phenomena, properly analyzed, could be understood if appropriate mathematical models could be constructed to describe their behaviors.

In summary, Paul A. Samuelson's very successful academic and research career was based on the following consequential set of conditions and circumstances:

- Samuelson came from a very distinguished line of elite academics who were involved in important research and scholarship, not just in economics, but also in the natural sciences and mathematics.
- Samuelson continued on this track and himself became the ancestor academic who produced students who tackled some of the most challenging problems and issues in economics.

- Samuelson attended and worked at two of the most elite universities in the world, namely, Harvard and MIT. Consequently, he had ready access to both resources and scientific networks of influence.
- Samuelson's and his students' work were awarded the highest honors in the field of economics. He and several of his students were given Nobel prizes for their contributions.

These factors were a direct result of Samuelson's stellar academic genealogy.

## CASE STUDY: RONALD E. MICKENS

Ronald E. Mickens is a mathematical scientist whose research centers on the modeling of systems. He has published almost 400 papers in peer-reviewed journals, and authored/co-authored twenty-two books. He grew up in Petersburg, Virginia, attending Peabody High School. After graduating from Fisk University, in 1964, he obtained his doctorate degree in theoretical physics in 1968 from Vanderbilt University and spent two years at the Center for Theoretical Physics at MIT as a National Science Foundation Fellow.

Mickens's research has been very broad in its subject matter and involved topics in areas such as (*Wikipedia*, 2025):

- high energy physics (asymptotic properties of scattering amplitudes),
- difference equations and their applications, and
- nonlinear oscillations in one-dimensional systems (Mickens, 2010).

Mickens's noteworthy accomplishments and contributions to mathematics and physics include (*Wikipedia*, 2025):

- the construction of nonstandard finite difference schemes (NFDS) for the discretization of differential equations (Mickens, 1994),
- construction and analysis of generalized trigonometric functions (Mickens, 2019), and
- writing of biographical essays on the lives, careers, and research of African American scientists.

Mickens was also among the first modern authors to write a book on difference equations and their applications (Mickens, 1987),

The AG charts for Mickens are given in Figures 4 and 5. Figure 4 is the AG which goes through his undergraduate advisor/mentor at Fisk University, Professor/Dr. James Raymond Lawson. Lawson received his BA in Physics at Fisk in 1935 and the Ph.D. from the University of Michigan in 1939. Likewise, Figure 5 is the AG that goes through Mickens's Ph.D. advisor/mentor, Professor/Dr. Wendell Gene Holladay. Holladay received his BA (1949) and MS (1950) degrees from Vanderbilt University, and his Ph.D. in Physics from the University of Wisconsin (1954).

A close inspection of Figure 4 shows that one of the connections for the first five levels of academic ancestry is that all of these scientists were involved in the area of infrared spectroscopy. It should also be pointed out that Mickens was heavily grounded in techniques and the methodology of IR-spectroscopy while an undergraduate at Fisk. Furthermore, all of the academic ancestors of Mickens were distinguished scientists who made important contributions to several areas of the physical sciences and scientific administration.

Likewise, Figure 5 includes the names of two individuals who received the highest award in physics, namely, the Nobel Prize. They are Max Born (1954 NP) and Maria Goeppert Mayer (1963 NP). This chart also contains the names of one of the foremost mathematicians of the 19th century, Karl Weirstrass, along with his student Carl Runge. Furthermore, this AG includes Friedrich Bessel and Carl Gauss, who are well known to students and professors in both physics and mathematics.

In summary, Ronald Mickens's AGs are populated by many of the 19th and 20th centuries' most accomplished scientists and mathematicians.

## CONCLUSION

This article has reported on certain aspects of academic genealogies (AG) as it applies (mainly) to scientists. We have not attempted to be complete, but only presented the essential features of how to actually construct an AG. Below we list several of the consequences which flow from having a knowledge of your or some other individual's AG:

- Having an AG chart allows a scientist to acquire a sense of their scientific heritage as it relates to their research.
- AGs of other scientists may be used to help in the selection of collaborators, which can then initiate the formation of an "invisible college."
- Knowledge of the successes and failures of researchers higher up on the AG trees can inform a scientist on paths to both take and to avoid, and thus eliminating previously made errors and mistakes.
- Knowledge of the AGs of collaborators can help in the recruitment of top level students and allow for the possibility of forming networks with elite scientists.
- Having an "outstanding" AG and membership in the appropriate invisible college enhances opportunities to be nominated and selected for top level awards and honors; it also increases the chances of receiving research funding; and brings the scientist to the attention of the scientific community to which they belong, thus increasing the possibility to be selected to positions of scientific leadership.
- An "outstanding" AG, along with distinguished research outputs, greatly increases the chances of obtaining a position at an elite university or research institute; consequently, in these positions they can reproduce their own research siblings.

The real significance of these six items becomes apparent when a comparison is made with the indicators that many use to gauge success in science:

- significant and impactful research;
- producing publications having high impact factors;
- receiving adequate funding for research and student support;
- having a tenured professorship, especially at a prestigious college, university, research institute, or government laboratory;
- collaborating with other highly successful scientists;
- producing successful students;
- being a leader in your discipline;
- holding positions of leadership in the appropriate professional societies;
- being invited to write a review on some technical issue of relevance to your discipline;
- being an active member and leader in your associated "invisible college"; and
- becoming a role model or inspiration for the next generation of scientists.

Finally, based on our experiences in gathering an assortment of materials to write this article, and discussions among ourselves and with others, we conclude that every academic, whether a scientist or not, should construct their own academic genealogical tree and use it to explore its various meanings, interpretations, and evaluations related to one's research and scholarly career, to date. This will prove to be a task that gives no regrets.

SELECTED ACADEMIC GENEALOGICAL WEBSITES

| | |
|---|---|
| Chemistry | https://web-genealogy.acs.illinois.edu |
| Mathematics | https://www.genealogy.math.ndsu |
| Philosophy | https://academictree.org/philosophy |
| Neuroscience | https://neurotree.org/neurtotree |
| Linguistics | https://academictree.org/linguistics |
| General | https://academictree.org/<br>(This one covers genealogies of more than 38 academic disciplines) |

Figure 1. Websites for genealogy searches for selected disciplines.

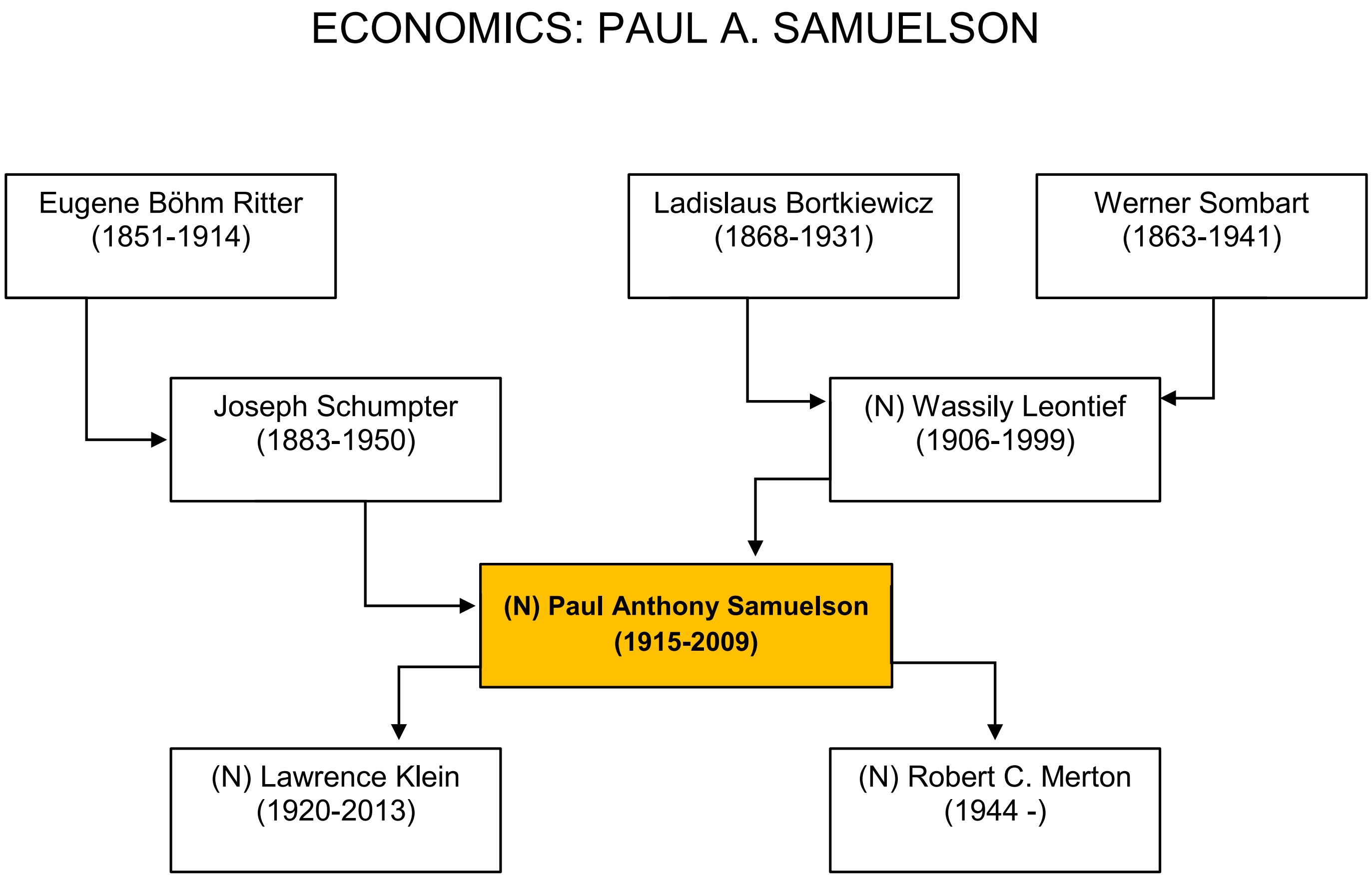

Figure 2. Truncated AG of academic connections for Paul A. Samuelson in the discipline of Economics. (N) denotes Nobel Prize recipient.

# MATHEMATICS/PHYSICS: PAUL A. SAMUELSON

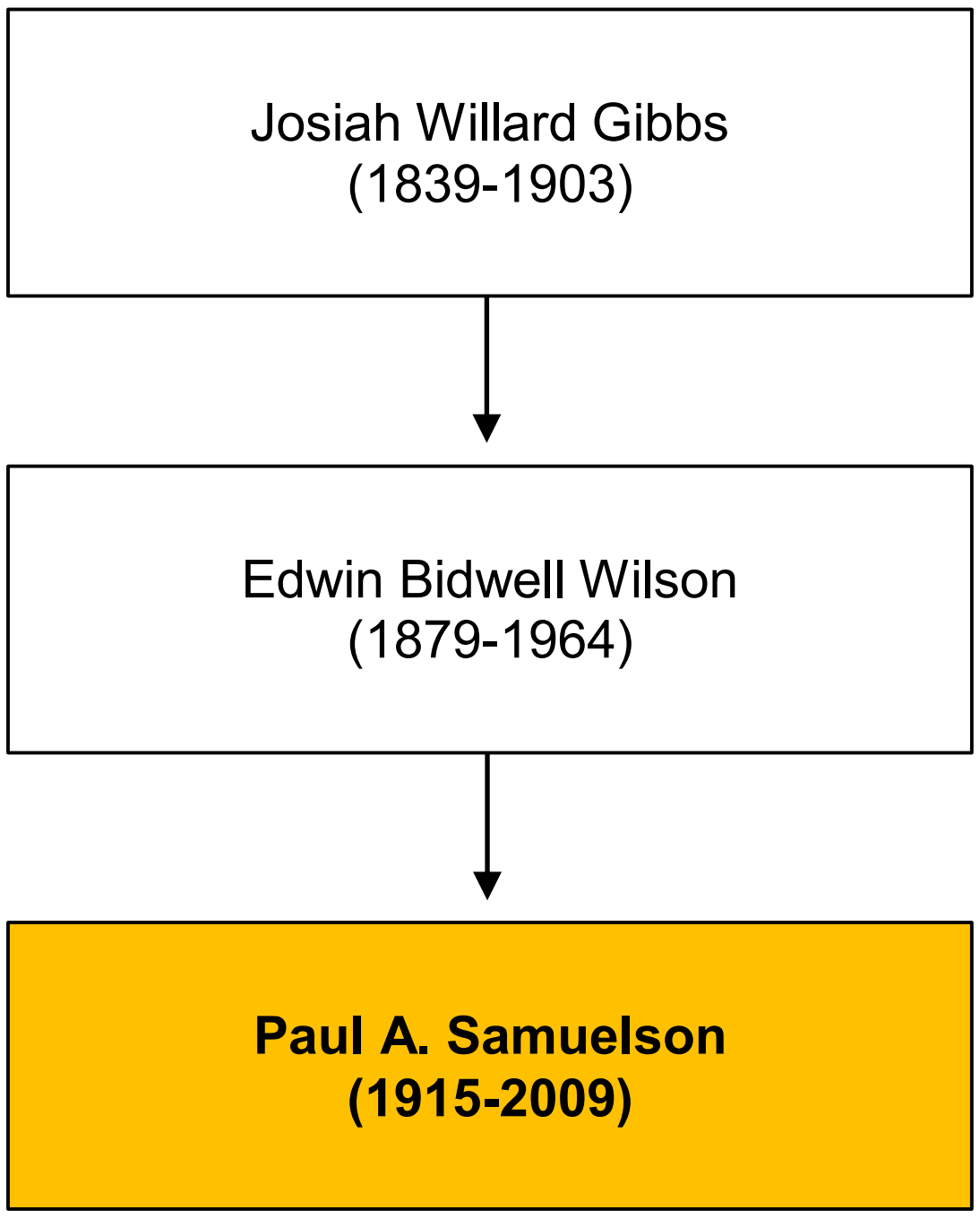

Figure 3. Paul A. Samuelson's academic connections to the discipline of mathematical/theoretical physics. Samuelson trasferred physical concepts to the field of economics using mathematics.

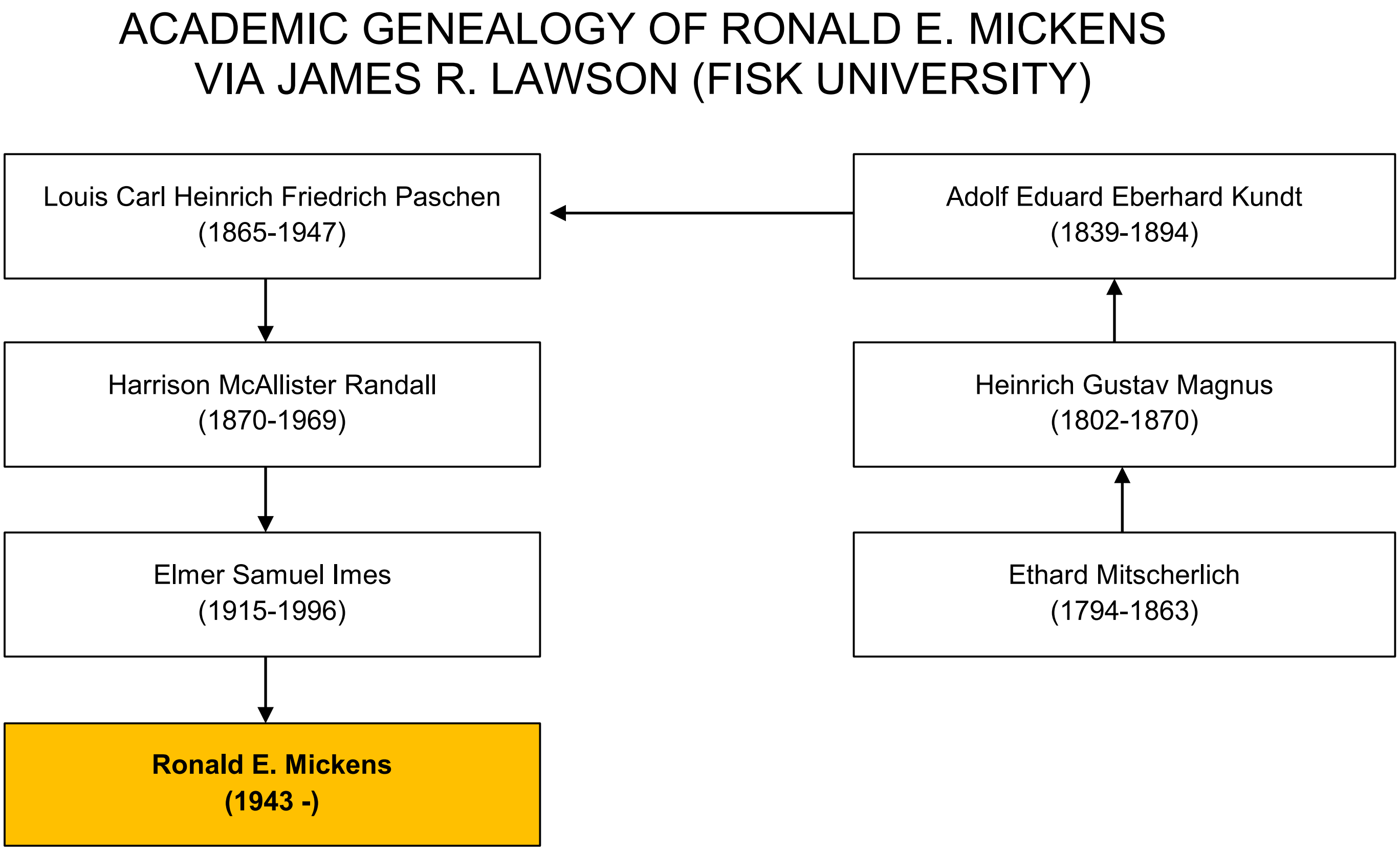

Figure 4. Academic genealogy of Ronald E. Mickens via his undergraduate mentor at Fisk University, James R. Lawson.

# ACADEMIC GENEALOGY OF RONALD E. MICKENS VIA WENDELL G. HOLLADAY (VANDERBILT UNIVERSITY)

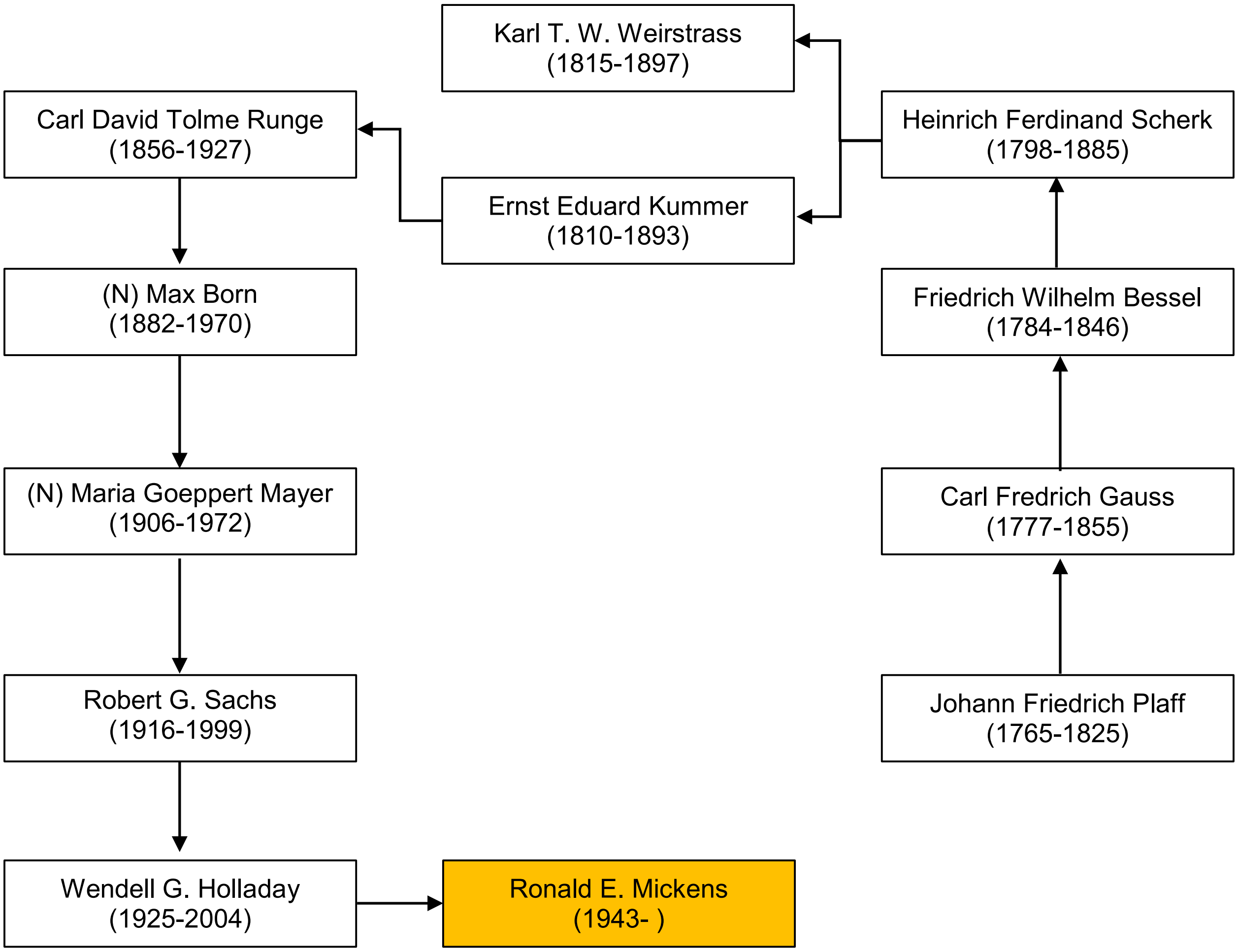

Figure 5. Academic genealogy of Ronald E. Mickens via his Ph.D. advisor, Wendell G. Holladay, at Vanderbilt University. (N) denotes Nobel Prize recipient.